\catcode`\@=11

\font\tenmsa=msam10
\font\sevenmsa=msam7
\font\fivemsa=msam5
\font\tenmsb=msbm10
\font\sevenmsb=msbm7
\font\fivemsb=msbm5
\newfam\msafam
\newfam\msbfam
\textfont\msafam=\tenmsa  \scriptfont\msafam=\sevenmsa
  \scriptscriptfont\msafam=\fivemsa
\textfont\msbfam=\tenmsb  \scriptfont\msbfam=\sevenmsb
  \scriptscriptfont\msbfam=\fivemsb

\def\hexnumber@#1{\ifnum#1<10 \number#1\else
 \ifnum#1=10 A\else\ifnum#1=11 B\else\ifnum#1=12 C\else
 \ifnum#1=13 D\else\ifnum#1=14 E\else\ifnum#1=15 F\fi\fi\fi\fi\fi\fi\fi}

\def\msa@{\hexnumber@\msafam}
\def\msb@{\hexnumber@\msbfam}
\mathchardef\boxdot="2\msa@00
\mathchardef\boxplus="2\msa@01
\mathchardef\boxtimes="2\msa@02
\mathchardef\square="0\msa@03
\mathchardef\blacksquare="0\msa@04
\mathchardef\centerdot="2\msa@05
\mathchardef\lozenge="0\msa@06
\mathchardef\blacklozenge="0\msa@07
\mathchardef\circlearrowright="3\msa@08
\mathchardef\circlearrowleft="3\msa@09
\mathchardef\rightleftharpoons="3\msa@0A
\mathchardef\leftrightharpoons="3\msa@0B
\mathchardef\boxminus="2\msa@0C
\mathchardef\Vdash="3\msa@0D
\mathchardef\Vvdash="3\msa@0E
\mathchardef\vDash="3\msa@0F
\mathchardef\twoheadrightarrow="3\msa@10
\mathchardef\twoheadleftarrow="3\msa@11
\mathchardef\leftleftarrows="3\msa@12
\mathchardef\rightrightarrows="3\msa@13
\mathchardef\upuparrows="3\msa@14
\mathchardef\downdownarrows="3\msa@15
\mathchardef\upharpoonright="3\msa@16

\mathchardef\downharpoonright="3\msa@17
\mathchardef\upharpoonleft="3\msa@18
\mathchardef\downharpoonleft="3\msa@19
\mathchardef\rightarrowtail="3\msa@1A
\mathchardef\leftarrowtail="3\msa@1B
\mathchardef\leftrightarrows="3\msa@1C
\mathchardef\rightleftarrows="3\msa@1D
\mathchardef\Lsh="3\msa@1E
\mathchardef\Rsh="3\msa@1F
\mathchardef\rightsquigarrow="3\msa@20
\mathchardef\leftrightsquigarrow="3\msa@21
\mathchardef\looparrowleft="3\msa@22
\mathchardef\looparrowright="3\msa@23
\mathchardef\circeq="3\msa@24
\mathchardef\succsim="3\msa@25
\mathchardef\gtrsim="3\msa@26
\mathchardef\gtrapprox="3\msa@27
\mathchardef\multimap="3\msa@28
\mathchardef\therefore="3\msa@29
\mathchardef\because="3\msa@2A
\mathchardef\doteqdot="3\msa@2B

\mathchardef\triangleq="3\msa@2C
\mathchardef\precsim="3\msa@2D
\mathchardef\lesssim="3\msa@2E
\mathchardef\lessapprox="3\msa@2F
\mathchardef\eqslantless="3\msa@30
\mathchardef\eqslantgtr="3\msa@31
\mathchardef\curlyeqprec="3\msa@32
\mathchardef\curlyeqsucc="3\msa@33
\mathchardef\preccurlyeq="3\msa@34
\mathchardef\leqq="3\msa@35
\mathchardef\leqslant="3\msa@36
\mathchardef\lessgtr="3\msa@37
\mathchardef\backprime="0\msa@38
\mathchardef\risingdotseq="3\msa@3A
\mathchardef\fallingdotseq="3\msa@3B
\mathchardef\succcurlyeq="3\msa@3C
\mathchardef\geqq="3\msa@3D
\mathchardef\geqslant="3\msa@3E
\mathchardef\gtrless="3\msa@3F
\mathchardef\sqsubset="3\msa@40
\mathchardef\sqsupset="3\msa@41
\mathchardef\trianglerighteq="3\msa@44
\mathchardef\trianglelefteq="3\msa@45
\mathchardef\bigstar="0\msa@46
\mathchardef\between="3\msa@47
\mathchardef\blacktriangledown="0\msa@48
\mathchardef\blacktriangleright="3\msa@49
\mathchardef\blacktriangleleft="3\msa@4A
\mathchardef\blacktriangle="0\msa@4E
\mathchardef\triangledown="0\msa@4F
\mathchardef\eqcirc="3\msa@50
\mathchardef\lesseqgtr="3\msa@51
\mathchardef\gtreqless="3\msa@52
\mathchardef\lesseqqgtr="3\msa@53
\mathchardef\gtreqqless="3\msa@54
\mathchardef\Rrightarrow="3\msa@56
\mathchardef\Lleftarrow="3\msa@57
\mathchardef\veebar="2\msa@59
\mathchardef\barwedge="2\msa@5A
\mathchardef\doublebarwedge="2\msa@5B
\mathchardef\angle="0\msa@5C
\mathchardef\measuredangle="0\msa@5D
\mathchardef\sphericalangle="0\msa@5E
\mathchardef\varpropto="3\msa@5F
\mathchardef\smallsmile="3\msa@60
\mathchardef\smallfrown="3\msa@61
\mathchardef\Subset="3\msa@62
\mathchardef\Supset="3\msa@63
\mathchardef\Cup="2\msa@64

\mathchardef\Cap="2\msa@65

\mathchardef\curlywedge="2\msa@66
\mathchardef\curlyvee="2\msa@67
\mathchardef\leftthreetimes="2\msa@68
\mathchardef\rightthreetimes="2\msa@69
\mathchardef\subseteqq="3\msa@6A
\mathchardef\supseteqq="3\msa@6B
\mathchardef\bumpeq="3\msa@6C
\mathchardef\Bumpeq="3\msa@6D
\mathchardef\lll="3\msa@6E

\mathchardef\ggg="3\msa@6F

\mathchardef\circledS="0\msa@73
\mathchardef\pitchfork="3\msa@74
\mathchardef\dotplus="2\msa@75
\mathchardef\backsim="3\msa@76
\mathchardef\backsimeq="3\msa@77
\mathchardef\complement="0\msa@7B
\mathchardef\intercal="2\msa@7C
\mathchardef\circledcirc="2\msa@7D
\mathchardef\circledast="2\msa@7E
\mathchardef\circleddash="2\msa@7F
\def\ulcorner{\delimiter"4\msa@70\msa@70 }
\def\urcorner{\delimiter"5\msa@71\msa@71 }
\def\llcorner{\delimiter"4\msa@78\msa@78 }
\def\lrcorner{\delimiter"5\msa@79\msa@79 }
\def\yen{\mathhexbox\msa@55 }
\def\checkmark{\mathhexbox\msa@58 }
\def\circledR{\mathhexbox\msa@72 }
\def\maltese{\mathhexbox\msa@7A }
\mathchardef\lvertneqq="3\msb@00
\mathchardef\gvertneqq="3\msb@01
\mathchardef\nleq="3\msb@02
\mathchardef\ngeq="3\msb@03
\mathchardef\nless="3\msb@04
\mathchardef\ngtr="3\msb@05
\mathchardef\nprec="3\msb@06
\mathchardef\nsucc="3\msb@07
\mathchardef\lneqq="3\msb@08
\mathchardef\gneqq="3\msb@09
\mathchardef\nleqslant="3\msb@0A
\mathchardef\ngeqslant="3\msb@0B
\mathchardef\lneq="3\msb@0C
\mathchardef\gneq="3\msb@0D
\mathchardef\npreceq="3\msb@0E
\mathchardef\nsucceq="3\msb@0F
\mathchardef\precnsim="3\msb@10
\mathchardef\succnsim="3\msb@11
\mathchardef\lnsim="3\msb@12
\mathchardef\gnsim="3\msb@13
\mathchardef\nleqq="3\msb@14
\mathchardef\ngeqq="3\msb@15
\mathchardef\precneqq="3\msb@16
\mathchardef\succneqq="3\msb@17
\mathchardef\precnapprox="3\msb@18
\mathchardef\succnapprox="3\msb@19
\mathchardef\lnapprox="3\msb@1A
\mathchardef\gnapprox="3\msb@1B
\mathchardef\nsim="3\msb@1C
\mathchardef\napprox="3\msb@1D
\mathchardef\nsubseteqq="3\msb@22
\mathchardef\nsupseteqq="3\msb@23
\mathchardef\subsetneqq="3\msb@24
\mathchardef\supsetneqq="3\msb@25
\mathchardef\subsetneq="3\msb@28
\mathchardef\supsetneq="3\msb@29
\mathchardef\nsubseteq="3\msb@2A
\mathchardef\nsupseteq="3\msb@2B
\mathchardef\nparallel="3\msb@2C
\mathchardef\nmid="3\msb@2D
\mathchardef\nshortmid="3\msb@2E
\mathchardef\nshortparallel="3\msb@2F
\mathchardef\nvdash="3\msb@30
\mathchardef\nVdash="3\msb@31
\mathchardef\nvDash="3\msb@32
\mathchardef\nVDash="3\msb@33
\mathchardef\ntrianglerighteq="3\msb@34
\mathchardef\ntrianglelefteq="3\msb@35
\mathchardef\ntriangleleft="3\msb@36
\mathchardef\ntriangleright="3\msb@37
\mathchardef\nleftarrow="3\msb@38
\mathchardef\nrightarrow="3\msb@39
\mathchardef\nLeftarrow="3\msb@3A
\mathchardef\nRightarrow="3\msb@3B
\mathchardef\nLeftrightarrow="3\msb@3C
\mathchardef\nleftrightarrow="3\msb@3D
\mathchardef\divideontimes="2\msb@3E
\mathchardef\varnothing="0\msb@3F
\mathchardef\nexists="0\msb@40
\mathchardef\mho="0\msb@66
\mathchardef\thorn="0\msb@67
\mathchardef\beth="0\msb@69
\mathchardef\gimel="0\msb@6A
\mathchardef\daleth="0\msb@6B
\mathchardef\lessdot="3\msb@6C
\mathchardef\gtrdot="3\msb@6D
\mathchardef\ltimes="2\msb@6E
\mathchardef\rtimes="2\msb@6F
\mathchardef\shortmid="3\msb@70
\mathchardef\shortparallel="3\msb@71
\mathchardef\smallsetminus="2\msb@72
\mathchardef\thicksim="3\msb@73
\mathchardef\thickapprox="3\msb@74
\mathchardef\approxeq="3\msb@75
\mathchardef\succapprox="3\msb@76
\mathchardef\precapprox="3\msb@77
\mathchardef\curvearrowleft="3\msb@78
\mathchardef\curvearrowright="3\msb@79
\mathchardef\digamma="0\msb@7A
\mathchardef\varkappa="0\msb@7B
\mathchardef\hslash="0\msb@7D
\mathchardef\hbar="0\msb@7E
\mathchardef\backepsilon="3\msb@7F
\def\Bbb{\ifmmode\let\next\Bbb@\else
 \def\next{\errmessage{Use \string\Bbb\space only in math mode}}\fi\next}
\def\Bbb@#1{{\Bbb@@{#1}}}
\def\Bbb@@#1{\fam\msbfam#1}

\catcode`\@=\active

\def\inv{^{\raise.15ex\hbox{${
  \scriptscriptstyle -}$}\kern-.05em 1}}

\def\Dsl{\,\raise.15ex\hbox{$/$}\mkern-13.5mu D}
\def\dsl{\raise.15ex\hbox{$/$}\kern-.57em\hbox{$\partial$}}

\def\lspace{\ifx\answ\bigans{}\else\qquad\fi}
\def\del{\partial}

\def\CR{\hbox{{$\cal R$}}}

\def\cg{\hbox{{\sl g}}} 

\def\lform{\hbox{$\sqcup$}\llap{\hbox{$\sqcap$}}}
\def\darr#1{\raise1.5ex\hbox{$\leftrightarrow$}
\mkern-16.5mu #1}

\def\h{{{1\over2}}}

\def\INT{{\textstyle \int\kern-.642em\int}}

\def\C{{\Bbb C}}
\def\Z{{\Bbb Z}}

\def\eps{{\epsilon}}

\def\rcross{{\triangleright\!\!\!<}}

\def\lcocross{{>\!\!\blacktriangleleft}}
\def\cobicross{{\triangleright\!\!\!\blacktriangleleft}}

\def\dcross{{\bowtie}}
\def\codcross{{\blacktriangleright\!\!\blacktriangleleft}}
\def\rbiprod{{\cdot\kern-.33em\triangleright\!\!\!<}}
\def\lbiprod{{>\!\!\!\triangleleft\kern-.33em\cdot}}

\def\tens{\mathop{\otimes}}

\def\la{{\triangleright}}\def\ra{{\triangleleft}}

\def\isom{{\cong}}

\def\id{{\rm id}}

\def\<{\langle}
\def\>{\rangle}

\def\emi#1{{\em #1\index{#1}}}

\def\vect{{\bf t}}\def\vecs{{\bf s}}
\def\vecu{{\bf u}}
\def\vecl{{\bf l}}

\def\<{\langle}
\def\>{\rangle}

\def\equad{\kern -1.7em}

\def\nquad{{\!\!\!\!\!\!}}

\def\o{{}_{\scriptscriptstyle(1)}}
\def\t{{}_{\scriptscriptstyle(2)}}
\def\th{{}_{\scriptscriptstyle(3)}}

\def\bo{{}^{\bar{\scriptscriptstyle(1)}}}
\def\bt{{}^{\bar{\scriptscriptstyle(2)}}}

\def\umo{{{}^{\scriptscriptstyle-(1)}}}
\def\umt{{{}^{\scriptscriptstyle-(2)}}}

\def\text#1{\mbox{\rm #1}}
\def\note#1{}

\def\blacksquare{{\lform}}
\def\frac#1#2{{{#1\over#2}}}
\def\from{{\leftarrow}}

\def\proof{\goodbreak\noindent{\bf Proof\quad}}

\def\endproof{{\ $\lform$}\bigskip }

\def\eqn#1#2{\begin{equation}#2\label{#1}\end{equation}}

\def\align#1{\begin{eqnarray*}#1\end{eqnarray*}}

\def\cmath#1{\[\begin{array}{c} #1 \end{array}\]}
\def\ceqn#1#2{\begin{equation}\label{#1}\begin{array}{c}#2\end{array}
\end{equation}}

\documentstyle[11pt]{article}
\newtheorem{lemma}{Lemma}[section] 
 \newtheorem{theorem}[lemma]{Theorem}

\begin{document}\baselineskip 20pt

{\ }\qquad\qquad  DAMTP/94-70 to app. Proc. 3rd Colloq. Quantum Groups, Prague,
June, 1994
\vspace{.2in}

\begin{center} {\LARGE SOME REMARKS ON THE QUANTUM DOUBLE}
\\ \baselineskip 13pt{\ }
{\ }\\ S. Majid\footnote{Royal Society University Research Fellow and Fellow of
Pembroke College, Cambridge}\\
{\ }\\
Department of Applied Mathematics \& Theoretical Physics\\
University of Cambridge, Cambridge CB3 9EW
\end{center}

\vspace{10pt}
\begin{quote}\baselineskip 13pt
\noindent{\bf Abstract}
 We recall the abstract theory of Hopf algebra bicrossproducts
and double cross products due to the author.  We use it to develop some
less-well known results about the quantum double as a twisting, as an extension
and as $q$-Lorentz group.


\end{quote}
\baselineskip 20pt

\section{Introduction}

The quantum double Hopf algebra $D(H)$ was introduced by Drinfeld\cite{Dri}.
Since then it has been extensively studied by the author in
\cite{Ma:phy}\cite{Ma:dou}\cite{Ma:ran}\cite{Ma:skl}\cite{Ma:poi}\cite{Ma:mec}
as well as by Reshetikhin and Semenov-Tsian Shanskii in \cite{ResSem:mat}.  It
was also proposed as $q$-Lorentz group in \cite{PodWor:def} and connected with
an R-matrix approach\cite{CWSSW:lor} in our paper \cite[Sec.  4]{Ma:poi}.  Here
we collect some of the main results and develop two less well known ones.  They
demonstrate the power of some algebraic techniques for Hopf algebras relating
to
cross products and cross coproducts.  We write formulae over $\C$ but in fact
these abstract constructions work over a general field too.  We use the
summation convention and also the standard abstract notation $\Delta h=h\o\tens
h\t$ for the quantum group coproduct.  This stands for a sum of terms in the
tensor product.

\section{Quantum double as a double cross product}

The definition of the quantum double that we use is the algebraic form due to
the author\cite{Ma:phy} as an example of a double cross product.  If $H$ is a
finite dimensional Hopf algebra or quantum group, its quantum double $D(H)$ is
built on the vector space $H^*\tens H$ with product \eqn{double}{ (a\tens
h)(b\tens g)=b\t a\tens h\t g \< Sh\o,b\o\> \< h\th,b\th\> } and tensor product
unit, counit and coproduct.  It is a quasitriangular Hopf algebra with
universal
R-matrix $\CR=(f^a\tens 1)\tens (1\tens e_a)$.  This is equivalent to
Drinfeld's
definition\cite{Dri} with generators and relations.

The first thing to notice is that the Hopf algebra construction here works
perfectly well if $H$ is not finite-dimensional:  in this case it is really a
function of {\em two} Hopf algebras $A,H$ which are dually paired by a map $\<\
,\ \>:H\tens A\to \C$.  We call this $D(A,H,\<\ ,\ \>)$ the {\em generalised
quantum double}.  Moreover, it is not necessary to have Hopf algebras either.
All this works at the bialgebra level if we replace $\<S(\ ),\ \>$ by the {\em
convolution inverse map} $\<\ ,\ \>^{-1}$.  The linear maps from $H\tens A\to
\C$ form an algebra using the coproduct of $H\tens A$, and we need our pairing
invertible in this algebra\cite{Ma:mor}.

Next, we introduce a general construction of which this generalised quantum
double is an example.  This is the concept of a {\em double cross product}.  We
consider two bialgebras or Hopf algebras which act on each other by maps \[
\ra:H\tens A\to H,\quad \la:H\tens A\to A\] which are coalgebra homomorphisms
(one says that $H$ is a right $A$-module coalgebra and $A$ is a left $H$-module
coalgebra).  We suppose further that these actions are a {\em matched pair} in
that they obey the compatibility conditions\cite{Ma:phy} \ceqn{dcross}{(hg)\ra
a=(h\ra (g\o\la a\o))(g\t\ra a\t),\quad 1\ra a=\eps(a)\\ h\la (ab)=(h\o\la
a\o)((h\t\ra a\t)\la b),\quad h\la 1=\eps(h)\\ h\o\ra a\o\tens h\t\la a\t=
h\t\ra a\t\tens h\o\la a\o} In this case, one has the theorem\cite{Ma:phy} that
there is a \emi{ double cross product bialgebra} $A\dcross H$ built on the
vector space $A\tens H$ with product \eqn{dcrossprod}{ (a\tens h)(b\tens g)=a
(h\o\la b\o)\tens (h\t\ra b\t) g} and tensor product unit, counit and
coproduct.
It is clear that $A,H$ are subalgebras and that $A\dcross H$ factorises into
them.  The theorem is:

\begin{theorem}cf.\cite{Ma:phy}.  If a bialgebra $X$ factorises in the sense
that there are sub-bialgebras \[ A{\buildrel i\over \hookrightarrow}
X{\buildrel
j\over \hookleftarrow} H\] such that $\cdot\circ(i\tens j):A\tens H\to X$ is an
isomorphism of vector spaces, then $A,H$ are a matched pair as above and
$X\isom
A\dcross H$.  \end{theorem} \proof This is based on \cite[Sec.  3.2]{Ma:phy}
but
in a cleaner form.  For this reason we give the full details.  Since
$\cdot\circ(i\tens j)$ is a linear isomorphism we have a well-defined linear
map
$\Psi:H\tens A\to A\tens H$ defined by \[ j(h)i(a)=\cdot\circ(i\tens j)\circ
\Psi(h\tens a).\] Associativity in $X$ and that $i,j$ are algebra maps tells us
that \align{ &&\equad \cdot\circ(i\tens j)\circ\Psi(hg\tens
a)=j(hg)i(a)=j(h)j(g)i(a)\\ &&=j(h)\cdot\circ(i\tens j)\circ\Psi(g\tens
a)=\cdot\circ(i\tens j)\circ (\id\tens \cdot)\circ
\Psi_{12}\circ\Psi_{23}(h\tens g\tens a)\\ &&\equad \cdot\circ(i\tens
j)\circ\Psi(h\tens ab)=j(h)i(ab)=j(h)i(a)i(b)\\ &&=\cdot\circ(i\tens
j)\circ\Psi(h\tens a)\, i(b)=\cdot\circ(i\tens j)\circ (\cdot\tens \id)\circ
\Psi_{23}\circ\Psi_{12}(h\tens a\tens b)} so we conclude from this and by a
similar consideration of the identity element in $X$ that
\ceqn{psifunct}{\Psi\circ(\cdot\tens \id)=(\id\tens
\cdot)\circ\Psi_{12}\circ\Psi_{23},\quad \Psi(1\tens a)=a\tens 1\\
\Psi\circ(\id\tens \cdot)=(\cdot\tens \id)\circ\Psi_{23}\circ\Psi_{12},\quad
\Psi(h\tens 1)=1\tens h} where the suffices refer to the tensor factor on which
$\Psi$ acts.  We see that the products in $H$ and $A$ respectively `commute'
with $\Psi$ in a certain sense.  This much works just at the algebra level
(factorisations of algebras are of this form).  Note that the braided tensor
product of algebras\cite{Ma:bra} is an example of such an algebra
factorisation.
Its introduction was motivated by this part of the double cross product
theorem.

Next, we use the counits $\eps$ to define linear maps $\ra:H\tens A\to H$ and
$\la:H\tens A\to A$ by \[\la=(\id\tens\eps)\circ\Psi,\quad
\ra=(\eps\tens\id)\circ\Psi.  \] Applying $\id\tens\eps$ to the first line of
(\ref{psifunct}) and $\eps\tens\id$ to the second tells us that $\la$ is a left
action and $\ra$ is a right action as required.  Applying $\eps\tens\id$ to the
first and $\id\tens\eps$ to the second tells us that \cmath{(hg)\ra
a=\cdot(h\ra\Psi(g\tens a)),\quad 1\ra a=\eps(a)\\ h\la(ab)=\cdot(\Psi(h\tens
a)\la b),\quad h\la 1=\eps(h).}  This much works for algebras equipped with
homomorphisms $\eps$.  Next we use that $i,j$ are coalgebra maps to deduce that
$\cdot\circ(i\tens j):A\tens H\to X$ and $\cdot\circ(j\tens i):H\tens A\to X$
are coalgebra maps too.  The inverse of $\cdot\circ(i\tens j)$ is also a
coalgebra map hence the ratio $\Psi$ is also a coalgebra map.  I.e., we
conclude
that \[ \Delta_{A\tens H}\circ\Psi=(\Psi\tens\Psi)\circ\Delta_{H\tens A},\quad
(\eps\tens\eps)\circ\Psi(h\tens a)=\eps(a)\eps(h).\] Now we apply
$\id\tens\eps\tens\eps\tens\id$ both sides of the first of these to conclude
that \[ h\o\la a\o\tens h\t\ra a\t=\Psi(h\tens a)\] which means that our
results
above prove the first two of (\ref{dcross}).  Applying instead the map
$\eps\tens\id\tens\id\tens\eps$ gives by contrast \[ h\o\ra a\o\tens h\t\la
a\t=\tau\circ \Psi(h\tens a)\] where $\tau$ is usual transposition.  This
proves
the third of (\ref{dcross}).  Likewise, applying instead
$\eps\tens\id\tens\eps\tens\id$ gives that $\ra$ is a coalgebra map, while
applying $\id\tens \eps\tens\id\tens\eps$ gives that $\la$ is a coalgebra map
too.  Hence we have all the conditions needed for a matched pair in the sense
of
(\ref{dcross}).  Finally, looking again at the equation for $j(h)i(a)$ above
tells us now that $\cdot(i\tens j)$ becomes an isomorphism between the
corresponding double cross product $A\dcross H$ and our original bialgebra $X$.
\endproof

The quantum double fits as an example of this theory as follows:  given two
paired bialgebra or Hopf algebras we define \[ h\ra a=h\t \< h\o,a\o\> ^{-1}\<
h\th,a\t\>,\quad h\la a=a\t \< h\o,a\o\> ^{-1}\< h\t,a\th\> \] and check that
they obey the conditions (\ref{dcross}) for a matched pair $(A^{\rm
op},H,\ra,\la)$.  We recover the generalised quantum double as \[ D(A,H,\<\ ,\
\>)=A^{\rm op}\dcross H\] which is how the generalised quantum double was first
introduced in \cite{Ma:mor}.  In the converse direction, one knows that the
double factorises\cite{Dri} and can use the above theorem to deduce $\ra,\la$.

Next, we make a trivial and completely cosmetic change:  we denote $A^{\rm op}$
in the above generalised quantum double by $A$.  Then of course this new $A$ is
no longer dually paired with $H$:  instead we denote the same linear map $\<\
,\
\>$ as a {\em skew pairing} $\sigma:H\tens A\to \C$.  It obeys the axioms
\eqn{skpair}{\nquad \sigma(hg\tens a)=\sigma(h\tens a\o)\sigma(g\tens a\t),\
\sigma(h\tens ab)=\sigma(h\o\tens b)\sigma(h\t\tens a).}  which are just the
axioms for $\<\ ,\ \>$ in terms of the new $A$ which has the opposite product
to
the old one which was dually paired.  Then \ceqn{gendou}{ D(A,H,\sigma)\equiv
A\dcross H\\ (a\tens h)(b\tens g)= a b\t\tens h\t g\, \sigma^{-1}(h\o\tens
b\o)\sigma(h\th\tens b\th)} with tensor product unit and counit is a double
cross product by \cmath{ h\ra a= h\t \sigma^{-1}(h\o\tens a\o)\sigma(h\th\tens
a\t)\\ h\la a= a\t \sigma^{-1}(h\o\tens a\o)\sigma(h\t\tens a\th).}

Finally, it is well-known in quantum groups that every construction has a dual
one.  It is an easy exercise to cast the above into their dual form (and not a
new result to do so!).  We list the result here for completeness.  Firstly, the
general construction is a {\em double cross coproduct}.  We consider two
bialgebras or Hopf algebras coacting on each other by maps \[ \alpha:A\to
A\tens
H,\quad \beta:H\to A\tens H\] which are algebra homomorphisms (comodule
algebras).  We require further that the coactions are matched in the sense
\ceqn{codcross}{ (\Delta\tens
\id)\circ\alpha(a)=\left((\id\tens\beta)\circ\alpha(a\o)\right)(1\tens\alpha(a\t))\\
(\id\tens\Delta)\circ\beta(h)=(\beta(h\o)\tens
1)\left((\alpha\tens\id)\circ\beta(h\t)\right)\\
\alpha(a)\beta(h)=\beta(h)\alpha(a).}  In this case, one has the theorem that
there is a bialgebra $H\codcross A$ with tensor product algebra structure and
counit, and \[ \Delta(h\tens a)=h\o\tens \alpha(a\o)\beta(h\t)\tens a\t.\] In
the Hopf algebra setting this is a Hopf algebra.  The double cross coproduct
comes equipped with bialgebra or Hopf algebra surjections
\eqn{codcrosspq}{H{\buildrel p\over\from} H\codcross A{\buildrel q\over\to}A}
into which factors the double cross coproduct decomposes by $(p\tens
q)\circ\Delta$.  Conversely, and bialgebra decomposing like this is a double
cross coproduct.

Our generalised quantum double example (\ref{gendou}) becomes a generalised
quantum codouble associated to a \emi{ skew copairing} between bialgebras or
Hopf algebras $A,H$.  This is an element $\sigma \in A\tens H$ such that
\eqn{skcopair}{ (\Delta\tens\id)\sigma =\sigma _{13}\sigma _{23},\qquad
(\id\tens\Delta)\sigma =\sigma _{13}\sigma _{12},} which is just the dual of
(\ref{skpair}).  We suppose $\sigma$ is invertible.  Then
\ceqn{gencodou}{D^*(H,A,\sigma)=H\codcross A\\ \Delta(h\tens a)=\sigma
_{23}^{-1}\Delta_{H\tens A}(h\tens a)\sigma _{23},\quad S(h\tens a)=\sigma
_{21}(Sh\tens Sa)\sigma _{21}^{-1}} with tensor product counit and algebra
structure is a double cross coproduct by coactions \[ \alpha(a)=\sigma
^{-1}(a\tens 1)\sigma ,\quad \beta(h)=\sigma ^{-1}(1\tens h)\sigma.\]

This general construction includes the dual of Drinfeld's quantum double as
$D(H)^*=H^{\rm cop}\codcross H^*$.  As well as being the dual of our
generalised
quantum double, the resulting examples (\ref{gencodou}) of double cross
coproducts associated to a skew copairing were found independently in
\cite{ResSem:mat} by another route as a `twisted product'.  The double cross
coproduct construction itself is however, much more general than this.

\section{Quantum double as a twisting, I}

If $A$ is a dual quasitriangular bialgebra (such as $A(R)$ according to results
in \cite[Sec.  3]{Ma:qua}) then the universal-R-matrix functional $\CR:A\tens
A\to\C$ as in \cite{Ma:bg}\cite{Ma:lin} obeys (\ref{skpair}).  Hence we have a
double cross product\cite{Ma:mor} \[ D(A,A,\CR)=A\dcross A\] which we
identified\cite{Ma:poi} as a natural construction for the $q$-Lorentz group.
It
is not in general isomorphic to Drinfeld's double, i.e.  we use the more
general
double (\ref{gendou}) coming out of the theory of double cross products.  There
is however, a Hopf algebra homomorphism \cite[Sec.  4]{Ma:poi} \[ A\dcross A\to
D^*(H),\quad a\tens b\mapsto (\id\tens a\o\tens b\o)(\CR^{-1}_{12}\CR_{31})
\tens a\t b\t\] if $A$ is dual to $H$.  It is an isomorphism in some
(factorisable) cases such as $SU_q(2)\dcross SU_q(2)$.  In matrix generators
$\vecs,\vect$ for the latter, the above map is\cite{Ma:poi} \[ \vecs\tens
1\mapsto S\vecl^+\tens\vect,\quad 1\tens\vect\mapsto S\vecl^-\tens\vect\] where
$\vecl^\pm$ are the matrix FRT generators\cite{FRT:lie} of $U_q(su_2)$.  Its
codouble as an algebra is the tensor product $U_q(su_2)\tens SU_q(2)$.  The
same
formula works for general $U_q(\cg)$.

As explained in \cite{Ma:poi} this is just a dual version of a corresponding
result in enveloping algebra form which is essentially due to
\cite{ResSem:mat}.
Thus, if $H$ is a strict quantum group with universal R-matrix (such as
$U_q(\cg)$) then $\sigma=\CR$ is a skew copairing and we have an example of a
generalised codouble or double cross coproduct \[ D^*(H,H,\CR)=H\codcross H\]
recovering the `twisted square' introduced in another framework in
\cite{ResSem:mat}.  These authors showed that \[ D(H)\to H\codcross H,\quad
a\tens h\mapsto \left((\id\tens a)(\CR_{31}^{-1}\CR_{23})\right)(\Delta h)\] is
a homomorphism of Hopf algebras, and argued that it should be an isomorphism if
$Q=\CR^{-1}\CR_{21}^{-1}$ is invertible when regarded as a linear map $H^*\to
H$
by evaluation on its first tensor factor (a version of the factorisable case).
The required inverse map was not given in \cite{ResSem:mat} and appears here I
think for the first time (being obtained in fact from the bosonisation theory
in
Section~6).  It is \[ h\tens g\mapsto (Q^{-1}(hSg\o))\o \<\CR\umt,
(Q^{-1}(hSg\o))\t\>\tens \CR\umo g\t.\] According to \cite[Sec.  4]{Ma:poi} the
example $U_q(su_2)\codcross U_q(su_2)$ is the correct construction for the
enveloping algebra of the q-Lorentz group dual to\cite{CWSSW:lor} and is
isomorphic to $D(U_q(su_2))$ by the above maps.  In matrix $\vecl^\pm$
generators the isomorphism from the quantum double to the twisted square is \[
\vect\tens 1\mapsto \vecl^-\tens\vecl^+,\quad 1\tens\vecl^\pm\mapsto
\vecl^\pm\tens\vecl^\pm.\]

We also introduced in \cite{Ma:poi} a $*$-structure in $A\dcross A$ given by
$(a\tens b)^*=b^*\tens a^*$ in the case when $\CR$ is of real-type in the sense
in \cite{Ma:mec}.  It is not a new result to repeat this in the dual form for
$H\codcross H$ where it immediately comes out as \[ (h\tens
g)^*=\CR_{21}(g^*\tens h^*)\CR_{21}^{-1};\quad \CR^{*\tens *}=\CR_{21}.\]

It is also clear from the form of the product of $A\dcross A$ and $H\codcross
H$
that they are examples of dual-twisting and twisting respectively in the more
modern sense by a non-Abelian 2-cocycle\cite{Ma:clau}, a theory due to Drinfeld
in \cite{Dri:qua} and restricted to Hopf algebras in \cite{GurMa:bra}.
Briefly,
a 2-cocycle {\em on} a quantum group is $\chi:A\tens A\to\C$ such that
\ceqn{2-cocycle-on}{\chi(b\o\tens c\o)\chi(a\tens b\t c\t)=\chi(a\o\tens
b\o)\chi(a\t b\t\tens c)\\ \chi(1\tens a)=\eps(a).}  The other side
$\chi(a\tens
1)=\eps(a)$ follows.  Given such a cocycle, the new product \eqn{cotwistprod}{
a\tilde\cdot b= \chi(a\o\tens b\o)a\t b\t \chi^{-1}(a\th\tens b\th)} defines a
new bialgebra $\widetilde{A}$.  Similarly for the antipode and any dual
quasitriangular structures of $\widetilde{A}$.  It is clear from the form
(\ref{gendou}) of $A\dcross A$ that it is the twisting by \[ \chi((a\tens
b)\tens (c\tens d))=\eps(a)\CR^{-1}(b\tens c)\eps(d)\] as a 2-cocycle on
$A\tens
A$.  In the braided geometrical approach to space time $SU_q(2)\tens SU_q(2)$
acts covariantly on a q-Euclidean space and this cocycle twists it to
$SU_q(2)\dcross SU_q(2)$ acting on q-Minkowski space, i.e.  it is physically a
quantum Wick rotation\cite{Ma:euc}.

Likewise, a 2-cocycle {\em for} a quantum group $H$ is $\chi\in H\tens H$ such
that\cite{Dri:qua}\cite{GurMa:bra} \eqn{2-cocycle}{ \chi_{23}(\id\tens\Delta)
\chi=\chi_{12}(\Delta\tens\id)\chi,\quad (\eps\tens\id)\chi=1} and means that
$\widetilde{H}$ with $\widetilde\Delta=\chi(\Delta\ )\chi^{-1}$ etc., is also a
quantum group.  This twisting is even more clear for $H\codcross H$ where the
coproduct (\ref{gencodou}) is obviously a twisting by a 2-cocycle
$\chi=\CR_{23}^{-1}$ for $H\tens H$.  So this picture of the q-Lorentz
enveloping algebra is likewise an immediate corollary of its description
introduced by the author in \cite{Ma:poi} as the twisted square
$U_q(su_2)\codcross U_q(su_2)$.  Some of this has subsequently been reiterated
by other authors following \cite{Ma:poi}.

\section{Quantum double as a twisting, II}

Now we look at a more well-known topic which is the quantum double of the Borel
subalgebra $U_q(b_-)$ say of $U_q(sl_2)$.  It is known that $D(U_q(b_+))$
projects onto $U_q(sl_2)$ and this is indeed how the quasitriangular structure
of the latter was deduced\cite{Dri}.  Likewise for general $U_q(\cg)$.

We begin by recalling this calculation with a modern proof based on q-geometry.
For $U_q(b_-)$ we take the Hopf algebra $H$ generated by $X,g,g^{-1}$ say and
\cmath{gX=q^2 Xg,\quad \Delta X=X\tens 1+g\tens X,\quad \Delta g=g\tens g\\
\eps
X=0,\quad \eps g=1 ,\quad Sg=g^{-1},\quad SX=-g^{-1}X.}  We show that when
$q^2\ne 0,1$ it is a self-dual Hopf algebra in the sense that it has a Hopf
algebra pairing with itself given by \[ \<g,g\>=q^2,\quad \<g,X\>=0,\quad
\<X,g\>=0,\quad \<X,X\>={1\over 1-q^{-2}}.\] This is essentially due to
Drinfeld
in \cite{Dri}, but here is a new proof.  We define a right action of $H$ on
itself by \[ \phi(X,g)\ra g= \phi(q^2X,q^2g),\quad \phi(X,g)\ra X=\lambda
{\del_q} \phi(X,g);\quad \lambda={1\over 1-q^{-2}}\] where $\phi(X,g)=\sum
\phi_{a b}X^ag^b$ is assumed to have powers of $X$ to the left and \[ \del_q
X^m
g^n=[m;q^2]X^{m-1}g^n;\quad [m;q^2]={1-q^{2m}\over1-q^2}\] is a q-derivative.
Since $\del_q$ lowers the degree by 1, it is clear that $(\phi\ra g)\ra X=q^2
(\phi\ra X)\ra g$ for all $\phi\in H$, so $\ra$ indeed defines a right action
of
our Hopf algebra on itself.  Moreover, one has \eqn{Tderiv}{ (\phi\psi)\ra
g=(\phi\ra g)(\psi\ra g),\quad (\phi\psi)\ra X=(\phi\ra X)\psi+(\phi\ra
g)(\psi\ra X)} for all normal-ordered polynomials $\phi(X,g),\psi(X,g)$.  Note
that the product $\phi\psi$ must be normal-ordered using the q-commutation
relations between $X,g$ before we can apply the definition of $\ra$.  The
action
of $g$ on products is clear since the normal ordering process commutes with
scaling of the generators by $q^2$.  The action of $X$ is easily verified on
monomials as \align{&&\equad (X^mg^nX^kg^l)\ra X=(q^{2nk}X^{m+k}g^{n+l})\ra
X=q^{2nk}\lambda[m+k;q^2] X^{m+k-1}g^{n+l}\\
&&=q^{2nk}\left(\lambda[m;q^2]+q^{2m}\lambda[k;q^2] \right) X^{m+k-1}g^{n+l}\\
&&=\lambda[m;q^2]X^{m-1}g^n X^k g^l+q^{2(m+n)}X^mg^n\lambda[k;q^2]X^k g^l\\
&&=((X^mg^n)\ra X)X^kg^l+((X^mg^n)\ra g)((X^kg^l)\ra X)} from the definition of
$\ra$ and the relations of $H$.  This result (\ref{Tderiv}) implies that the
product of $H$ is covariant (a module algebra) under the action $\ra$.

Given this action $\ra$, we now define \[ \<h,a\>=\eps(h\ra a)\] and deduce
that
it obeys half the axioms of a pairing from our module algebra property
(\ref{Tderiv}).  But since the resulting pairing \eqn{selfpair}{
\<X^mg^n,X^kg^l\>=\eps\left(((X^kg^l)\ra X^m)\ra
g^n\right)=\delta_{m,k}\lambda^m [m;q^2]!  q^{2nl}} is symmetric in the two
arguments of $\<\ ,\ \>$, we also deduce the other half by symmetry.

This proof also has the merit that the nondegeneracy of the pairing is clear,
at
least when $q$ is generic.  This is because
$\<\phi(X,g),X^mg^n\>=\lambda^m(\del_q^m\phi)(0,q^{2n})$ vanishing for all
$m,n$
implies that $\phi$ vanishes too.  Reducing non-degeneracy to the q-derivative
or higher-dimensional braided-derivative\cite{Ma:fre} appears to be a useful
strategy of proof which I have not seen elsewhere.

It is now an easy matter to compute $D(U_q(b_-))=H^{\rm op}\dcross H$ as
generated by $H=U_q(b_-)$ as above and $H^{\rm op}=U_q(b_+)$ with generators
$\bar X,\bar g,\bar g^{-1}$ say.  The result from (\ref{double}) is
\ceqn{Dsl2relns}{ gX=q^2 Xg,\quad g\bar X=q^{-2} \bar X g,\quad X\bar
X-q^{-2}\bar X X={\bar g g-1\over q^2-1}\nonumber\\ \Delta X=X\tens 1+g\tens
X,\quad \Delta \bar X=\bar X\tens 1+\bar g\tens \bar X} with the corresponding
counit and antipode.  The generators $g,\bar g$ are group-like.

We want to make two comments about this quantum double.  The first working over
formal powerseries $\C[[t]]$ in a deformation parameter $t$, and the second in
Section~5 working over $\C$.  In the first case we change variables \[
X=q^{-{H\over 2}}X_-,\quad \bar X=q^{-{H\over 2}}X_+,\quad g=q^{-H}q^{C},\quad
\bar g =q^{-H}q^{{-C}}\] to new generators $X_\pm, H,C$ then we get the usual
algebra relations of $U_q(sl_2)$ along with $C$ central and primitive and \[
\Delta X_\pm=X_\pm\tens q^{H\over 2}+q^{-{H\over 2}}q^{\mp C}\tens X_\pm.\]
Following Drinfeld one can now derive the universal R-matrix of $U_q(sl_2)$
from
the one of $D(U_q(b_-))$ after projecting by $C=0$.

On the other hand, we can consider the 1-dimensional quantum group
$U_{q^{-1}}(\C)$ with generator $C$ and \[ \Delta C=C\tens 1+1\tens C,\quad
\CR=q^{-\h C\tens C}\] This is called the quantum line in\cite{LyuMa:fou}.  On
the Hopf algebra $U_q(sl_2)\tens U_{q^{-1}}(\C)$ we define \[ \chi=q^{-\h
C\tens
H}\] and check that it is a 2-cocycle.  Then $[C\tens H,(\ )\tens X_\pm]=\pm 2C
(\ )\tens X_\pm$ tells us that the coproducts of $X_\pm$ in the double are just
the conjugation by $\chi$ of the usual coproduct of $U_q(sl_2)$, i.e.  we have
\[ D(U_q(sl_2))\isom \widetilde{ U_q(sl_2)\tens U_{q^{-1}}(\C)}\] as twisted by
this cocycle.  With a bit more work we have this as a twisting of
quasitriangular Hopf algebras.

\section{Quantum double as a cocycle bicrossproduct}

Now we recall another aspect of abstract Hopf algebra theory, which is the
notion of a cocycle cross product.  Thus a (right) {\em cocycle action} of a
Hopf algebra $H$ on an algebra $A$ is linear maps \[ \ra:A\tens H\to A,\quad
\chi:H\tens H\to A\] where \ceqn{cy-rmodalg}{1\ra h=1\eps(h),\quad (ab)\ra h=
(a\ra h\o)(b\ra h\t)\\ a\ra 1=a,\quad \chi(h\o\tens g\o)((a\ra h\t)\ra g\t)=
(a\ra (h\o g\o)) \chi(h\t\tens g\t)\\ \chi(h\o g\o\tens f\o)(\chi(h\t\tens
g\t)\ra f\t)=\chi(h\tens g\o f\o)\chi(g\t\tens f\t)\\ \chi(1\tens
h)=\chi(h\tens
1)=\eps(h)} for all $h,g,f\in H$ and $a,b\in A$.  In this situation, there is a
right cocycle cross product algebra $H\rcross_\chi A$ on the vector space
$H\tens A$ with product \eqn{cyrcross}{ (h\tens a)(g\tens b)= h\o
g\o\tens\chi(h\t\tens g\t)(a\ra g\th) b.}  That this product is associative
follows from the assumptions (\ref{cy-rmodalg}).  A cohomological picture of
such cocycle cross coproducts is due to \cite{Doi:equ}.  They can be
interpreted
as trivial quantum principal bundles in quantum group gauge theory
\cite{BrzMa:gau}, with $\chi$ a new quantum number not present classically.

In the dual setting we have of course the notion of a cocycle coaction.  So a
left cocycle coaction of a Hopf algebra $A$ on a coalgebra $H$ for example
means
maps \[ \beta:H\to A\tens H,\quad \psi:H\to A\tens A\] obeying some axioms dual
to those above.  See \cite{Ma:mor}.  We will not need the most general form
with
the dual-cocycle $\psi$.  In this case we have a usual cross coproduct:  if
$\beta(h)=h\bo\tens h\bt$ is a coaction of $A$ on the coalgebra $H$ and
respects
its structure in the sense \eqn{comodcoalg}{\nquad
(\id\tens\Delta)\circ\beta(h)=h\o\bo h\t\bo\tens h\o\bt\tens h\t\bt,\quad
(\id\tens\eps)\circ\beta=\eps} (a comodule coalgebra) then there is a cross
coalgebra $H\lcocross A$ built on $H\tens A$ with \eqn{lcocross}{\nquad
\Delta(h\tens a)=h\o\tens h\t\bo a\o\tens h\t\bt\tens a\t,\quad \eps(h\tens
a)=\eps(h)\eps(a).}

Putting these ideas together we suppose next that $H$ acts on $A$ by a cocycle
action as above and $A$ coacts back on $H$.  We can demand conditions on these
such that the resulting cross product algebra and coalgebra fit together to
form
a {\em bicrossproduct Hopf algebra} $H\cobicross A$.  This was introduced for
general Hopf algebras by the author in \cite{Ma:phy} (without cocycles) and
\cite{Ma:mor} (with cocycles), with some examples in \cite{Ma:phy}\cite{Ma:pla}
and \cite{MaSoi:bic} respectively.  We refer to these papers for the precise
compatibility conditions which ensure that the bicrossproduct is a Hopf
algebra.
A general feature of the bicrossproduct is that \[ A\hookrightarrow H\cobicross
A\twoheadrightarrow H\] by the obvious inclusion and the obvious projection via
$\eps$.  There is a theorem that every extension of Hopf algebras like this
which is invertible in a certain sense (the extension should be cleft and
cocleft) is of this bicrossproduct type, possibly with cocycles.

For a recent application of this theory of extensions to physics, see
\cite{MaRue:bic}.  We want to note only that that quantum double of
$H=U_q(b_-)$
in the form computed in the last section, is exactly such an extension \[
\C\Z\hookrightarrow D(U_q(b_-)){\buildrel p\over \twoheadrightarrow}
U_q(sl_2).\] Here $p$ is defined by \eqn{projDsl2}{ \nquad p(X)=F\equiv
q^{-{H\over 2}}X_-,\quad p(\bar X)= E\equiv q^{-{H\over 2}}X_+,\quad
p(g)=p(\bar
g)=q^{-H}} where the $ E,F,q^{\pm H}$ can be considered as generating a version
of $U_q(sl_2)$.  It is a sub-Hopf algebra of the usual form since we do not use
$q^{\pm {H\over 2}}$ themselves.  With this form of $U_q(sl_2)$ understood, we
have $p$ as a surjection.  Next, it is clear that the invertible element
$K=\bar
g g^{-1}$ generates $\C\Z\subset D(H)$ as a commutative sub-Hopf algebra.
Moreover, it is clear that $D(H)\isom U_q(sl_2)\tens \C\Z$ as vector spaces by
$X^m\bar X^n g^k K^l\mapsto F^m E^n q^{-Hk}\tens K^l$.  Finally, one verifies
that this isomorphism respects multiplication by $K$ from the right, and the
coaction of $U_q(sl_2)$ from the left in the manner required for a Hopf algebra
extension.

This and some technical cleftness conditions which one can also verify means
that our quantum double must be a cocycle bicrossproduct.  It is \[\nquad\nquad
D(U_q(b_-))\isom U_q(sl_2)\cobicross_\chi \C\Z,\quad \beta(\cases{ E&\cr F&\cr
q^{-H}&}\nquad)=\cases{K\tens E&\cr 1\tens F&\cr 1\tens q^{-H}}\nquad,\ \chi(
E\tens F)={1-K\over 1-q^{-2}}\] with $\chi=\eps\tens\eps$ on the other
generators of $U_q(sl_2)$.  The action $\ra$ in the bicrossproduct and the
dual-cocycle $\psi$ are both trivial.  We identify $X=F\tens 1$, $\bar X=
E\tens
1$, $g=q^{-H}\tens 1$ and $K\equiv 1\tens K$.  Then the product from
(\ref{cyrcross}) as modified by the cocycle $\chi$ gives \align{&&\bar X X=(
E\tens 1)(F\tens 1)= E\o F\o\tens \chi( E\t\tens F\t)\\ &&= EF\tens 1+ g^2\tens
\chi( E\tens F)= EF\tens 1+g^2\tens {1-K\over 1-q^{-2}}\\ &&=q^2F E\tens 1+
{1-g^2\over1-q^{-2}}\tens 1+g^2\tens {1-K\over 1-q^{-2}}\\ &&=q^2(F\tens 1)(
E\tens 1)+{1-g^2\tens K\over 1-q^{-2}}=q^2 X\bar X+{1-g^2 K\over 1-q^{-2}}}
which are the relations (\ref{Dsl2relns}) for the quantum double.  The
coalgebra
structure (\ref{lcocross}) as modified by $\beta$ is \[ \Delta \bar X=\Delta(
E\tens 1)= E\tens 1\tens 1\tens 1+q^{-H}\tens \beta( E)\tens 1=\bar X\tens
1+gK\tens \bar X\] as required in (\ref{Dsl2relns}).  This completes our
bicrossproduct description of $D(U_q(b_-))$.  We see that as an algebra it is a
central extension of $U_q(sl_2)$ by the cocycle $\chi$, and as a coalgebra it
is
a cross coproduct.  Since $\C\Z=\C(S^1)$ we can think of this as a quantum
principle bundle over $S^1$ with fibre $SU_q(2)^\star$ in the geometrical
picture of \cite{BrzMa:gau}.  For a full geometrical picture one should
consider
the $*$-structure also, with the above as the chiral part of the full picture.

\section{Quantum double as a bosonisation}

For completeness, we conclude here with a brief mention of another abstract
result on the quantum double already published in
\cite{Ma:dou}\cite{Ma:skl}\cite{Ma:mec}.  This applies to the case when $A$ is
a
dual quasitriangular Hopf algebra or bialgebra (such as $A(R)$).  According to
the theory of transmutation\cite{Ma:bra} it has a braided group version $B$
which lives in the braided category of representations of the corresponding
quasitriangular Hopf algebra $H$ (such as $U_q(\cg)$).  For example, the
transmutation of the quantum matrices $A(R)$ is just the braided matrices
$B(R)$
introduced in this way in \cite{Ma:exa} with matrix generators $\vecu$ and
relations $R_{21}\vecu_1R\vecu_2=\vecu_2R_{21}\vecu_1R$.  It turns quantum
$2\times 2$ matrices $M_q(2)$ into braided (hermitian) $2\times 2$ matrices
$BM_q(2)$ or $q$-Minkowski space.  Likewise at the Hopf algebra level it turns
$SU_q(2)$ into the braided group $BSU_q(2)$ which is the mass-shell in
$q$-Minkowski space\cite{Ma:mec}.

On the other hand, there is a general theory of {\em bosonisation}
\cite{Ma:bos}
which turns any braided group $B$ in such a representation category into an
ordinary quantum group $B\lbiprod H$.  For the algebra we make a cross product
as in (\ref{cyrcross}) but with a left action (say) rather than a right one,
and
without a cocycle.  By definition, $B$ is covariant under $H$ and we use this
action $\la$ for the cross product.  For the coproduct we use the induced left
coaction\cite{Ma:dou} \[ \beta:B\to H\tens B,\quad \beta(b)=\CR_{21}\la b\] and
make the cross coproduct (\ref{lcocross}).  This is a generalisation to braided
groups of the Jordan-Wigner transformation for superalgebras.

When applied to braided planes, this bosonisation gives us $q$-Poincar\'e Hopf
algebras\cite{Ma:poi}.  On the other hand, when applied to the above-mentioned
braided group obtained from $A$ dual to $H$, it gives the quantum
double\cite{Ma:dou}\cite{Ma:skl} \[ B\lbiprod H\isom D(H).\] Cross product
algebras can also be considered as quantisation of position functions $B$ by
momentum quantum group $H$, which is the interpretation developed in
\cite{Ma:mec}.  We can also consider the quantum double in this form as a
trivial quantum principal bundle\cite{BrzMa:gau}.  For example,
$BSU_q(2)\lbiprod U_q(su_2)$ is a bundle over the mass-shell in $q$-Minkowski
space with fibre $SU_q(2)^\star$.

\end{document}